# Using Google Analytics to Support Cybersecurity Forensics

Han Qin
Central Analytics
Navigant Consulting, Inc.
Washington, D.C. USA
Email: han.qin@navigant.com

Kit Riehle
Legal Technology Solutions
Navigant Consulting, Inc.
Washington, D.C. USA
Email: kit.riehle@navigant.com

Haozhen Zhao
Central Analytics
Navigant Consulting, Inc.
Washington, D.C. USA
Email: haozhen.zhao@navigant.com

*Abstract*—Web traffic is a valuable data source, typically used in the marketing space to track brand awareness and advertising effectiveness. However, web traffic is also a rich source of information for cybersecurity monitoring efforts. To better understand the threat of malicious cyber actors, this study develops a methodology to monitor and evaluate web activity using data archived from Google Analytics. Google Analytics collects and aggregates web traffic, including information about web visitors' location, date and time of visit, visited webpages, and searched keywords. This study seeks to streamline analysis of this data and uses rule-based anomaly detection and predictive modeling to identify web traffic that deviates from normal patterns. Rather than evaluating pieces of web traffic individually, the methodology seeks to emulate real user behavior by creating a new unit of analysis: the user session. User sessions group individual pieces of traffic from the same location and date, which transforms the available information from single point-in-time snapshots to dynamic sessions showing users' trajectory and intent. The result is faster and better insight into large volumes of noisy web traffic.

## I. Introduction

The internet provides companies with a valuable platform to promote brand awareness and connect with current and potential customers. However, a web presence also makes companies vulnerable to malicious cyber actors and theft of proprietary information. The cyberspace is particularly difficult to secure for a variety of reasons, including malicious actors' ability to operate anonymously from anywhere in the world, the interconnectedness of companies' physical and digital systems, and the ever-evolving nature of cyber networks [1]. Typical cyber-attacks include phishing, ransomware, theft, insertion of spyware or malware, distributed denial of service (DDoS) attacks, and drive-by downloads. As a result, companies are forced to spend vast human and financial resources in attempt to keep pace with cyber threats that are constantly evolving.

In this climate, cyber monitoring offers a valuable tool for companies to track engagement with their online content. By analyzing web traffic, companies can better understand and anticipate future threats, empowering them to move from a reactive to proactive position.

This paper demonstrates how companies can benefit from both the descriptive and predictive analysis of web traffic data. In the descriptive phase, web traffic is aggregated to support continuous monitoring and trend identification, which helps define a baseline of normal user engagement. In the predictive phase, web traffic is analyzed to identify activity that deviates from this baseline. Specifically, this study applies rule-based anomaly detection and predictive modeling to label web activity as 'normal' or 'abnormal'. The work of this paper demonstrates how analytics can cut through a company's noisy web traffic to identify key trends and abnormalities. By applying this approach, companies can learn what information is being targeted on their websites and by whom, thus arming them to take protective measures.

### A. Web Traffic Data

Web traffic data is critical for cybersecurity monitoring efforts as it provides a record of every action performed by a visitor on a website. Typically, this information is compiled from web services' log files and may be accessed via third-party platforms, such as Google Analytics or Adobe Analytics.

This study uses real-world web traffic from a large, multinational company. The data spans almost 10 years and contains over 50 million web traffic records, each representing a visit to one of the company's webpages. The traffic includes visits to the company's job postings, product and service descriptions, and employee biographies, among other things. The data was archived by Google Analytics and accessed via the Google Analytics Reporting Application Programming Interface (API). It includes information about visitors' country, city, date of visit, time of visit, duration of visit, visited webpages, and searched keywords.

Searched keywords are particularly valuable for cybersecurity monitoring efforts as they provide insight into visitors' motivation for engaging with a website. Whereas visited webpages provide a high-level outline of a visitor's path ('where' the visitor went), searched keywords show what information the visitor sought ('why' they visited a webpage). Queried phrases reveal not only what a visitor found, but also what a visitor hoped to find and could not. In this study, information about visited webpages and searched keywords is combined to analyze both visitor behavior and intent. This is particularly valuable for web activity that is deemed 'unusual' or malicious, as it can help to profile cyber thieves and their targets.

## II. MOTIVATION

This study focuses on the methods used to perform cybersecurity monitoring. The methodology seeks to identify patterns in web activity and uses rule-based anomaly detection and predictive analysis to classify activity patterns as benign or suspicious. The monitoring is designed such that the definition of 'suspicious activity' can evolve over time and can be customized by company. A company, for example, might define suspicious activity as targeted research of a proprietary product, whereas a government might define it as repeated searches for employees with sensitive specializations. In general, this study demonstrates how analytics can support companies' cyber monitoring efforts and alleviate some of the burden on cyber analysts.

Predictive modeling is already recognized as a valuable method for working with unstructured text [2]. In legal and regulatory matters, for example, predictive modeling can automate the review of documents and target text that is relevant to a case, thus saving attorneys significant time and cost. Similarly, predictive modeling offers a promising tool for the cybersecurity space, where many of the same data challenges apply. Like attorneys, cyber analysts must cut through large volumes of data and be adept at working with text, namely website URLs and searched keywords.

Traditionally, however, web traffic has presented an additional challenge that has prevented this data from being analyzed in the same manner as documents: the unit of analysis. Whereas a document presents a long, ordered series of words, a piece of web traffic contains only a webpage URL and, if provided, a few searched keywords. The highly standardized nature of webpage URLs reduces the dimensionality in this type of text. A cyber analyst reviewing web traffic might find that the only meaningful information in a URL is a few words after the generic 'http://www.' and domain name.

To address this challenge, the study develops a new unit of analysis: the user session. The user session groups individual pieces of web traffic from the same location and approximate time with the logic that these records likely represent activity from the same visitor. User sessions mimic documents in that they combine snippets of text to form ordered series, much like arranging sentences to form a paragraph. Doing so transforms the available web traffic data from single point-in-time snapshots to dynamic sessions showing users' trajectory, from their starting webpage through their ending webpage.

The benefit of user sessions is two-fold. Conceptually, measuring sessions rather than individual clicks offers a more nuanced measure of user activity. Two users might both visit webpages with sensitive content, but, by examining their user sessions, it is possible to discern the user who visited accidently versus the user who reached the webpage methodically. Practically, user sessions are beneficial because the longer unit supports more data features, which predictive models use to achieve better accuracy and predictive power.

## III. DATA MINING

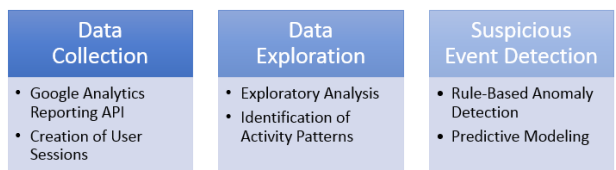

Figure 1. Methodology Phases

The process of data mining was divided into three phases: data collection, data exploration, and suspicious event detection, which included both rule-based anomaly detection and predictive modeling. The phases build to offer a historical view ('what has happened?'), a descriptive view ('what is happening?'), and finally a prescriptive view ('what requires monitoring?').

### A. Data Collection

The data for this study is real-world web traffic from a large, multinational company and includes any visit to the company's website while the company was actively registered with Google Analytics. The data spans almost 10 years and contains over 50 million web traffic records, each representing a visit to one of the company's webpages. The data was archived by Google Analytics and accessed via the Google Analytics Reporting Application Programming Interface (API).

Once collected, the data was stored in a Python data structure for ease of processing and analysis. The individual pieces of web traffic were grouped by location and approximate time to form 'user sessions' with the logic that these records likely represent activity from the same visitor. A minimum of three pieces of web traffic were required to form a session. Within each session, the web traffic was ordered chronologically, thus allowing the session to capture a user's trajectory from webpage to webpage. Records with unidentified locations (e.g., traffic for which the city or country is "Unknown") were excluded from the session aggregation.

### B. Data Exploration

Data exploration was performed on both the individual pieces of web traffic and aggregated user sessions. Statistics were calculated to identify traffic volume by a variety of data fields (see Table 1).

This phase of analysis helped to identify the baseline of activity for the studied company. Exploring traffic by location, for example, revealed that the studied company receives an average of 12.5% of its traffic from outside of the United States. Of foreign traffic, the most common countries of origination are the United Kingdom, India, and Germany.

By studying these trends, the study began to identify web activity that deviates from the baseline. One week in March 2015, for example, was noted for exhibiting an unusually large volume of traffic from Japan. Examination of these records revealed that many of the visited webpages and searched keywords relate to a thought-piece released by the company days prior and focused on Japan.

Table 1. Traffic Volume Statistics

| Level of Aggregation | Example |
|---|---|
| **Visit date** | |
| Daily | 1/1/2015 |
| Weekly | Week of 1/1/2015 |
| Monthly | January 2015 |
| Yearly | 2015 |
| **Visit time of day** | |
| Minute | 12:01am 1/1/2015 |
| Hourly | 12am 1/1/2015 |
| **Visitor location** | |
| City | London |
| Country | United Kingdom |
| Region | Europe |
| **Visited webpage** | |
| Webpage | http://www.navigant.com/careers/search-jobs |
| Directory webpage | http://www.navigant.com/careers |
| **Searched keywords** | |
| Keywords | 'Quarterly report' |
| Keyword category | Financials |

In other instances, the 'abnormal' web activity was less innocuous. The analysis revealed attempts to access employee-only resource webpages from countries where the studied company does not operate, including Syria, Libya, and Ukraine. Similarly, the analysis found targeted searches of employees' biographies and contact information from these same countries. These events were often observed in clusters, over the span of a few days, and motivated the study to develop more robust methods to detect suspicious activity.

*C. Suspicious Event Detection*

The discoveries of the data exploration phase helped to profile when activity is suspicious or warrants further review. The data exploration phase found that web activity was easiest to classify as benign or suspicious when evaluating user sessions rather than individual pieces of traffic. The sessions have the benefit of showing users' trajectory, from their starting webpage through their ending webpage, which helps to differentiate between accidental clicks versus purposeful searches.

Using user sessions, the study developed two methods to programmatically detect suspicious events. The first method was rule-based anomaly detection and the second method was predictive modeling, which included two implementations of supervised learning models.

*1) Rule-Based Anomaly Detection*

This method sought to define a set of rules that determine when a user session passes from benign to suspicious. The rules were developed based on observations made in the data exploration phase (see Table 2).

Table 2. Initial Rules for Anomaly Detection

| **Traffic from non-business dealing countries with searches for:** |
|---|
| Sensitive products |
| Sensitive service lines |
| Employee contact information |
| Employee biographies |
| |
| **Attempts to visit employee-only resource pages from:** |
| Countries without current business operations |
| One location repeatedly |
| |
| **Traffic that exceeds a threshold volume in a given time period** |

This phase was performed iteratively such that results from one round could inform the development of additional rules. For example, results from the initial set of rules revealed that suspicious events often include visits to the studied company's newsletter registration page and enquiry form to request more information. This discovery informed the creation of additional rules (see Table 3).

Of the 50,578 user sessions (see Table 4) in the data, rule-based anomaly detection identified 8% or 4,041 sessions as suspicious. Manual review of these results confirmed that a majority of the 4,041 sessions were suspicious. However, the results were also found to contain an overrepresentation of benign activity patterns, such as targeted research of employees paired with visits to the company's job application page. Rule-based anomaly detection was thus found to have high recall, but low precision.

*2) Predictive Modeling*

This method expanded on the rule-based approach to provide more refined anomaly detection through machine learning. The study tested the effectiveness of two predictive models: a support vector machine and a logistic regression. A training set for the models was created by combining 11,980 benign sessions with 766 suspicious sessions. The 766 suspicious sessions were selected from the rule-based results and manually confirmed to represent potentially malicious behavior. A test set was created from the remaining 37,832 sessions.

*a) Support Vector Machine*

A support vector machine was developed to classify sessions as benign or suspicious. The model used the sessions' country and city of origin, visited URLs, and searched keywords as features. The model was run with five-fold cross-validation and found to yield 95% accuracy in predicting suspicious sessions.

*b) Logistic Regression*

The study also tested the effectiveness of logistic regression. Whereas the support vector machine gave each session a binary classification ('suspicious' or 'not'), the logistic regression predicted the likelihood that each session was suspicious and therefore provided more nuanced results. This model was also run with five-fold cross-validation and yielded an average accuracy of 96%. Thus, the logistic regression was found to be slightly more effective in this particular use case.

Table 3. Additional Rules for Anomaly Detection

| **Traffic from non-business dealing countries with visits to:** |
|---|
| Newsletter registration page |
| Enquiry form |

Table 4. Creation of User Sessions

| **Sessions Created** | 50,578 |
|---|---|
| **Pieces of Traffic per Session** | |
| Minimum | 3 |
| Average | 9.1 |

## IV. POTENTIAL INDUSTRY USES

### A. Cybersecurity in the Legal Space

While cybersecurity is critical for industries such as banking and retail, the legal space is another industry that needs to take heed. Law firms and legal service providers frequently handle high-stakes cases and harbor sensitive information, making them vulnerable to phishing attempts. The nature of the legal industry requires law firms and legal service providers to have direct access to their clients' data, which can include personal correspondence, intellectual property, financial statements, health records, and trade secrets.

Proprietary data is particularly attractive to cyber thieves as it can offer a competitive advantage and can be easily monetized. Cyber theft of information about a pending business deal, for example, can allow thieves to profit from the insider knowledge. In March 2016, a Russian hacker targeted 48 law firms, specially seeking to obtain and sell confidential information about pending mergers and acquisitions [3].

Faced with this threat, the American Bar Association has developed cybersecurity standards for law firms. The ABA Cybersecurity Legal Task Force enforces the standards and holds law firms accountable for the loss of any client data. "Cybersecurity no longer can be relegated to the IT department or be part of general guidelines on computer use. Cybersecurity, or as some industries call it, cyberrisk, is part of doing business" [4].

In this environment, a streamlined cybersecurity approach like the one developed by this study could offer substantial benefit at low cost. Web traffic is a readily-available and low-cost source of information that can provide direct insight into a law firm's online presence. By monitoring web traffic, a firm could learn how their data is being targeted and by whom, thus arming them to take protective measures

## V. CONCLUSION AND FUTURE WORK

### A. Current Approach

This study demonstrates that analytics can cut through a company's noisy web traffic to identify key trends and abnormalities. The study develops a methodology that uses descriptive and predictive analysis of web traffic to reveal important insights about user engagement, including detecting potentially malicious activity. By applying rule-based anomaly detection and predictive modeling, the study differentiates between benign and suspicious web activity with a high rate of accuracy. This approach offers an accessible and inexpensive tool for companies to proactively monitor their cyber-presence, which is critical in the today's business operating environment.

### B. Future Work

The lessons-learned from this study suggest the following ideas for future work:
- Incorporate additional information about user engagement with a company and its employees, including communications through email, social media, and LinkedIn
- Assess additional machine learning algorithms and ensemble methods, in which machine learning algorithms are combined to maximize predictive performance
- Leverage advanced technologies, such as Hadoop, to support faster analysis of large amounts of web traffic data

REFERENCES

[1] Cybersecurity Overview. Department of Homeland Security, 27 Sept. 2016, www.dhs.gov/cybersecurity-overview.

[2] Chhatwal, R., Huber-Fliflet, N., Keeling, R., Zhang, J., & Zhao, H. (2016). Empirical evaluations of preprocessing parameters' impact on predictive coding's effectiveness. 2016 IEEE International Conference on Big Data (Big Data). doi:10.1109/bigdata.2016.7840747.

[3] Evans, J. Randolph, and Shari L. Klevens. Cybersecurity: You Can't Afford to Ignore It Anymore. American Bar Association, 25 Apr. 2016.

[4] Sobowale, Julie. "Law Firms Must Manage Cybersecurity Risks." ABA Journal, American Bar Association, Mar. 2017, www.abajournal.com/magazine/article/managing_cybersecurity_risk.